\documentclass[reprint, prl, showkeys]{revtex4-2}
\usepackage{hyperref}

\usepackage[utf8]{inputenc}
\usepackage[english]{babel}
\usepackage{amsmath}
\usepackage{latexsym}
\usepackage{graphicx}
\usepackage{float}
\usepackage{multirow}
\usepackage{amssymb}
\usepackage{mathrsfs}
\usepackage{makecell}
\usepackage{rotating}

\begin{document}
\title{Magnetization steps of $J_{1}$ quintets in the hcp lattice. }

\author{X. Gratens and V. Chitta}
\affiliation{Instituto de Física, Universidade de São Paulo, 05315-970, São Paulo, Brazil}
\begin{abstract}
The magnetization steps of quintet clusters, consisting of five identical magnetic ions coupled by isotropic nearest neighbors antiferromagnetic exchange interaction in the hcp lattice, have been investigated. In that model, called the $J_1$ model, there are 17 types of quintets. The values of the magnetic field of the magnetization steps of the clusters have been determined by the numerical diagonalization of the spin Hamiltonian. The simulations have been performed for quintets composed of individual spin $S$ = 1/2, 3/2 and 5/2. The contribution of the quintets to  the effective concentration (or technical saturation) as a function of the magnetic ion concentration was also calculated.  
\end{abstract}

\maketitle

\section{I. Introduction}
The method of magnetization steps (MSTs) is a 20-year-old technique that provides a direct measure of the exchange constant for antiferromagnetic (AF) magnetic clusters \cite{10.1063/1.1507808}. The method has been successfully applied to diluted magnetic semiconductors (DMS) with Rock Salt \cite{PhysRevB.57.7854,XGratens_2000,GRATENS20001519}, Zinc Blende \cite{PhysRevLett.80.5425}, and Wurtzite crystal structures \cite{PhysRevB.69.125209, bib2}. Another application of the MST method is to obtain information about the spatial distribution of the magnetic ions over all the cation lattice. This can be performed by comparing the technical saturation value of the magnetization \cite{PhysRevB.30.4021} to predictions given by cluster models. The method was recently applied to DMS nanoparticles \cite{C8CP02870B}. It was found that the magnetic ions are distributed with a higher concentration  near the surface of the nanoparticles.

The present work is focused on the MSTs of AF quintets, which consist of five identical magnetic ions, within the hcp cation lattice. We have used a simple cluster model based on only one isotropic AF exchange interaction between nearest neighbors and the Zeeman energy. In the next this model is called $J_{1}$ model. To simplify the calculation, the distinction between the nearest neighbors out of the \textbf{c}-plane ($J_{1}^{out}$) of the wurtzite structure and the nearest neighbors in the same \textbf{c}-plane ($J_{1}^{in}$) \cite{PhysRevB.69.125209} is not made here. The model includes the contribution of the singles (\textit{S}), pairs (\textit{P}), two types of triplets, open triplet (\textit{OT}) and closed triplet (\textit{CT}), and six types of quartets (tetrahedron (\textit{Td}), square (\textit{Sq}), double triangle (\textit{DT}), funnel (\textit{F}), propeller (\textit{Pr}) and string (\textit{St})) \cite{10.1063/1.1507808}. Based on the cluster tables given in the supplementary files of Ref.\cite{10.1063/1.1507808}, we have seventeen types of quintets in the $J_{1}$ model. The fields at the MSTs of the quintets with individual spin \textit{S} = 1/2, 3/2 and 5/2 have been calculated. The expression of the ratio of the technical saturation value to the true saturation value of the magnetization has been determined including the contributions of the clusters up to the quintets.   

\section{II. Quintet Types}

In the $J_{1}$ model with nonequivalent nearest neighbors exchange interactions $J_{1}^{in} \neq J_{1}^{out}$, we have 90 types of quintets in the hcp lattice 90 \cite{ClusterTables}. Assuming in the following that $J_{1}^{in} = J_{1}^{out} = J_{1}$, the 90 types of quintets can be grouped into 17 types. Figure 1 displays one possible configuration of quintet cluster for each of the 17 types. In order of the number of exchange interaction, we have three types with 4 exchange interactions, 4 types with 5 exchange interactions, 4 types with 6 exchange interactions, 3 types with 7 exchange interactions, 2 types with 8 exchange interactions and finally only one type with 9 exchange interactions. Four of these seventeen quintet types also exist in the body center cubic lattice. The results for these four types (V.1, V.2, V.3 and V.7 in Figure 1) have already been published for $S = 5/2$ \cite{PhysRevB.64.214424}.

\begin{figure}
\centering
\includegraphics[width=9.00 cm]{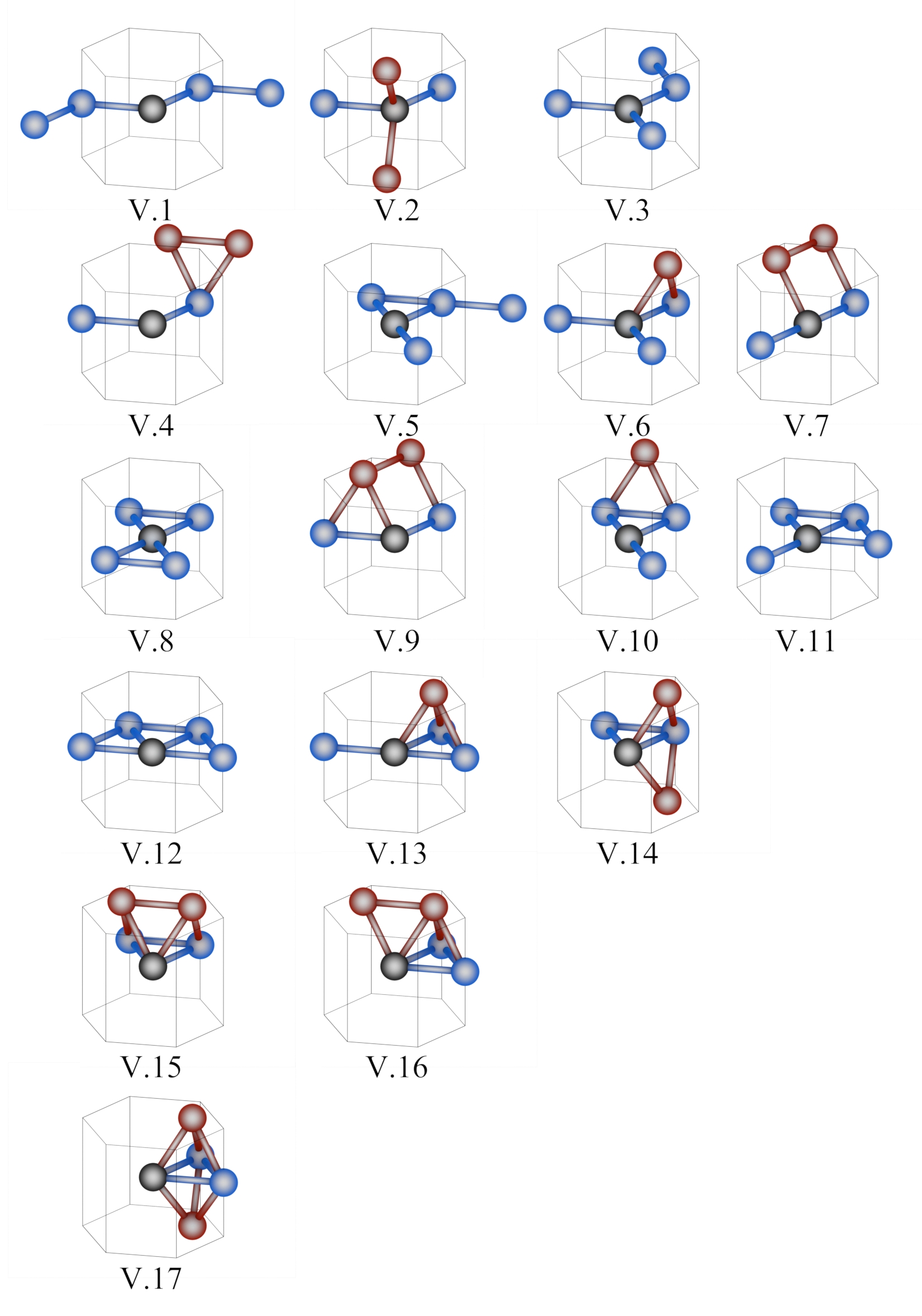}
\caption{The seventeen types of quintets in the $J_{1}$ model for hcp lattice. For each type, one possible configuration of the cluster is displayed. The clusters are grouped into lines based on their exchange bond number. Clusters in the first, second, third, fourth, and fifth lines have 4, 5, 6, 7, 8, and 9 bonds, respectively. The atom in black represents the central atom. In blue are the atoms located in the \textbf{c}-plane of the central atom while the atoms in red are located in a different \textbf{c}-plane.}
\label{Figure_1}
\end{figure} 

\section{III. Magnetization steps} 

The general form of the spin Hamiltonian of five spins coupled by an isotropic exchange interaction $J$ and in presence of magnetic field is:

\begin{equation}
\mathscr{H} = g \mu _{B} \sum \limits_{i=1}^{5} H  S_{iz} -2\sum \limits_{k = 1}^{4} \sum \limits_{l = k+1}^{5}J_{kl} \textbf{S}_{\textbf{k}}\cdot \textbf{S}_{\textbf{l}}
\label{Equation_1}
\end{equation}
where $J_{kl}$ = \textit{J} or = 0.
 
The first term represent the Zeeman interaction with isotropic \textit{g}-factor and the magnetic field \textbf{H} along the quantization axis \textit{z}. %The sum is performed for all atoms composing the quintet. 
The second term represents the isotropic exchange interaction and it is the sum over all spin pairs (\textit{k},\textit{l}) involving \textit{J}.

As example, the Hamiltonian for the "\textit{string}" quintet (labeled V.1 in Figure 1) is given by: 

\begin{equation} \label{Equation_2}
\begin{split}
\mathscr{H} = g \mu _{B} H (S_{1z}+S_{2z}+S_{3z}+S_{4z}+S_{5z})\\
-2J_{1}(\textbf{S}_{\textbf{1}}\cdot \textbf{S}_{\textbf{2}}+\textbf{S}_{\textbf{2}}\cdot \textbf{S}_{\textbf{3}}+\textbf{S}_{\textbf{3}}\cdot \textbf{S}_{\textbf{4}}+\textbf{S}_{\textbf{4}}\cdot \textbf{S}_{\textbf{5}})
\end{split}
\end{equation}

The Hamiltonian matrix of Eq.(1) with dimension ($(2S+1)^{5}$ $\times$ $(2S+1)^{5}$) is constructed using the basis of states $\left| m_{S_{1}}=l, m_{S_{2}}=k, m_{S_{3}}=n, m_{S_{4}}=p, m_{S_{5}}=q \right\rangle $ = $\left| l, k, n, p, q\right\rangle $ with \textit{l}, \textit{k}, \textit{n}, \textit{p}, \textit{q} = -$S$, -$S$+1, ..., $S$-1, $S$. 
The energy levels have been calculated through the numerical diagonalization of the Hamiltonian matrix of Eq.(1). The fields at the MSTs that correspond to the energy level crossings, which lead to a change of the ground state, have been determined by the bisection numerical method. The obtained normalized fields $\alpha= g \mu_{B}H /|2 J/k_{B}|$ at the MSTs for the seventeen quintet types composed of individual spin \textit{S} = 1/2, 3/2 or 5/2 are given in Table I. The ground state, $S_{T}= S_{1}+S_{2}+S_{3}+S_{4}+S_{5}$, at zero magnetic field is also given in the table.         

The magnetization vs magnetic field trace of the cluster can be calculated, using the energy levels obtained by numerical diagonalisation of the Hamiltonian matrix of Eq.(1), the  partition function \textit{Z}, and the magnetic free energy \textit{F}:

\begin{equation} \label{Equation_3}
\begin{split}
Z = \sum\limits_{i}  exp(-E_{i} / k_{B} T)
\end{split}
\end{equation}

\begin{equation} \label{Equation_4}
\begin{split}
F = -k_{\mathrm{B}}T~ln(Z)
\end{split}
\end{equation}
by:

\begin{equation} \label{Equation_5}
\begin{split}
M = (\partial F / \partial H)_{T}
\end{split}
\end{equation}

When the calculation is performed, numerical overflow problems may arise at very low temperatures or with high magnetic fields. They are avoided by zeroing the energy of the ground state for each value of the magnetic field:\\ 

\begin{equation} \label{Equation_6}
\begin{split}
F = -k_{\mathrm{B}} T\Big(~ln \Big[\sum \limits_{i}  exp(-(E_{i}-E_{0}) / k_{\mathrm{B}} T)\Big]-E_{0}/ k_{\mathrm{B}} T\Big)
\end{split}
\end{equation}
 
\begin{figure}
\centering
\includegraphics[width=9.00 cm]{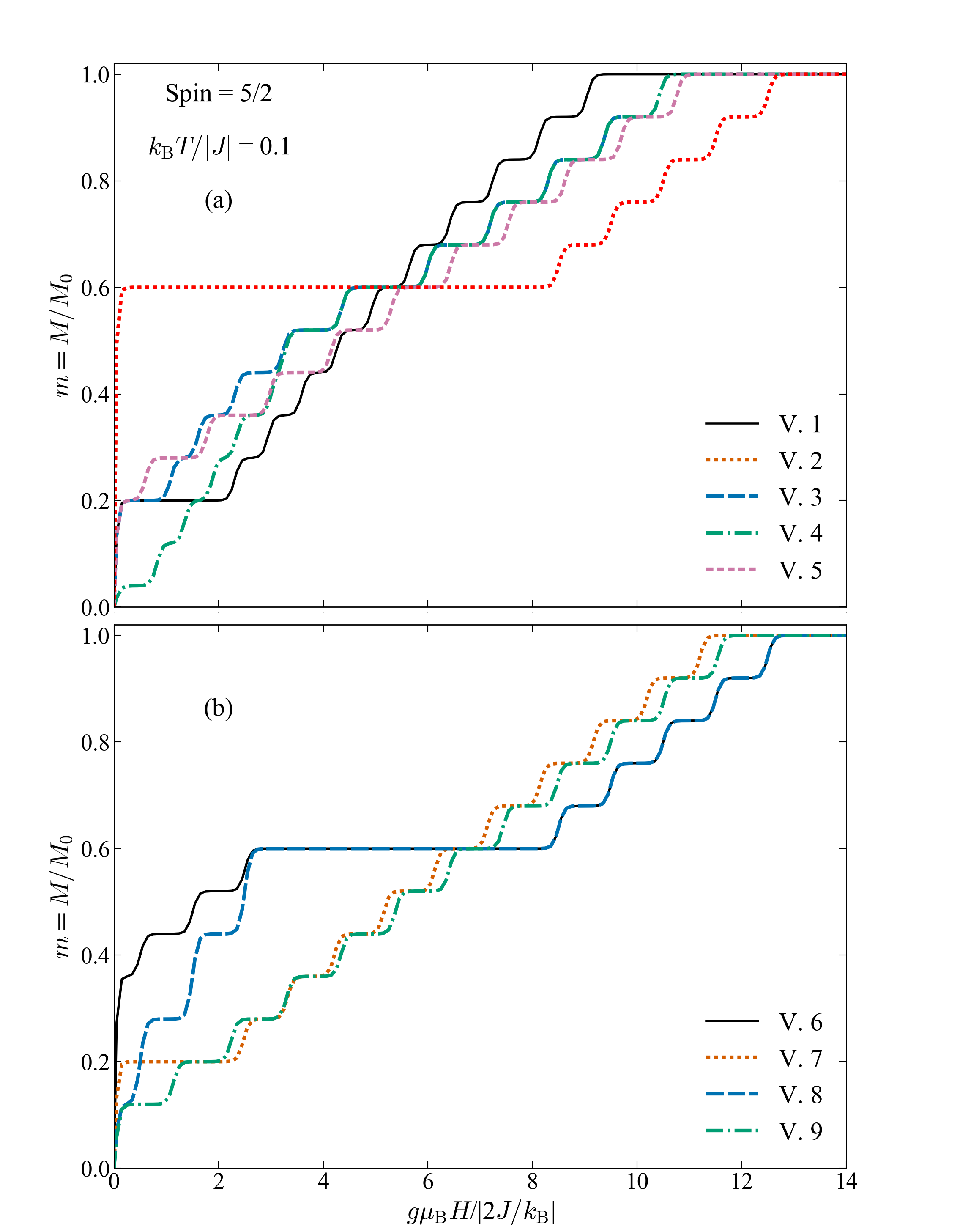}
\caption{Calculated magnetization curves of quintets (V.1 to V.9 of Figure 1), composed of individual spin \textit{S} = 5/2. The magnetization has been normalized to its saturation magnetization value. The calculation was made for $k_{\mathrm{B}}T/|J|$ = 0.1.}
\label{Figure1}
\end{figure}

\begin{figure}
\centering
\includegraphics[width=9.00 cm]{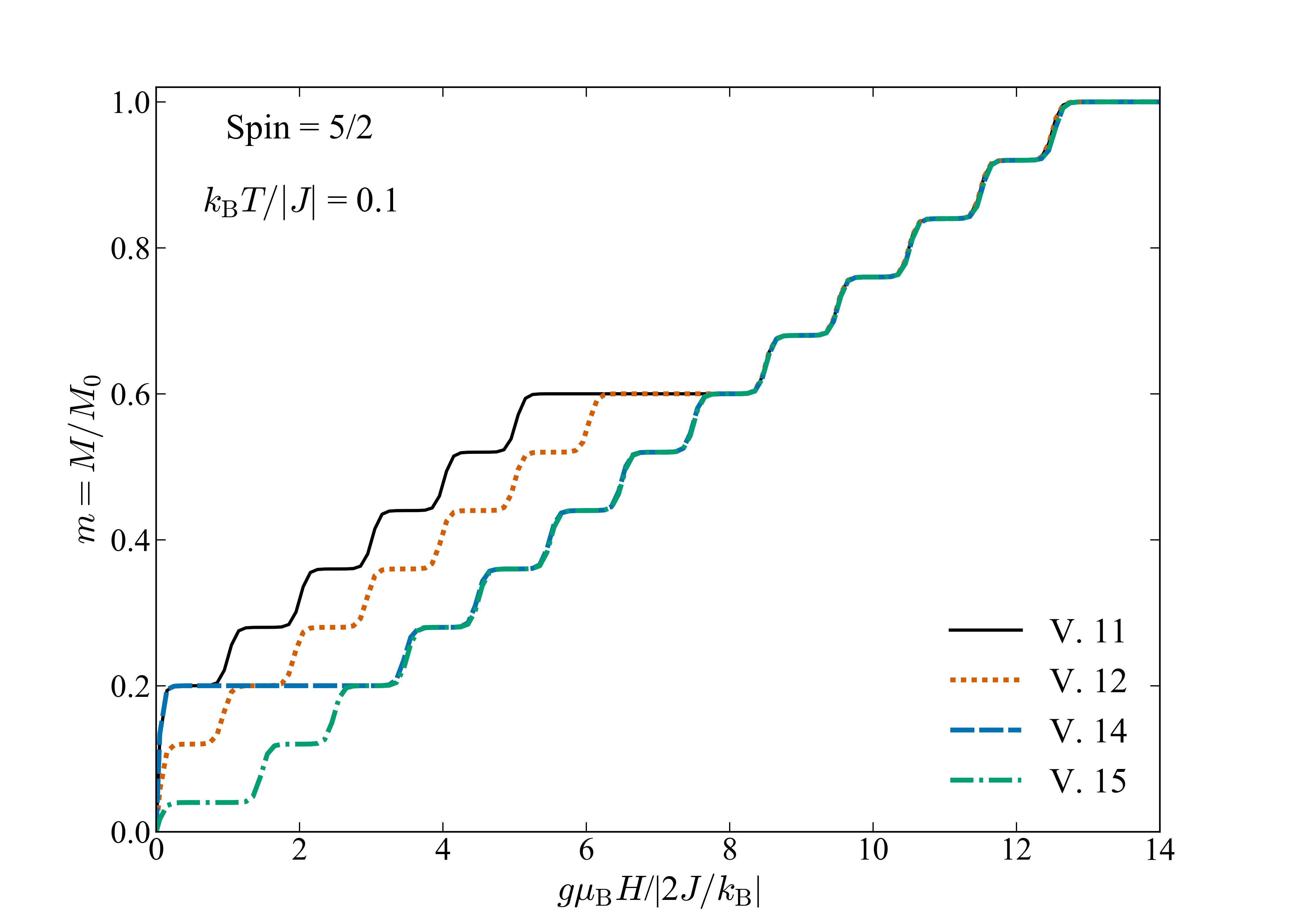}
\caption{Theoretical normalized magnetization curves of quintets V.11, V.12, V.14 and V.15 composed of individual spin \textit{S} = 5/2. The traces were calculated for $k_{\mathrm{B}}T/|J|$ = 0.1.}
\label{Figure1}
\end{figure}

Figure 2 and 3 show the calculated magnetization traces of quintets composed of individual spin \textit{S} = 5/2. The magnetization has been normalized to its saturation value $M_{0}$ = 5 $g \mu_{B}S$. The traces were calculated for $k_{B}T/|J|$ = 0.1. The magnetization curves for the quintets V.10, V.13, V.16 and V.17 are not displayed. They are very similar to the traces obtained for V.7, V.11, V.14 and V.15 respectively. The quintet type with the lower number of MSTs is V.2 with a total of 5 MSTs. The quintets V.6 and V.8 have seven and eight MSTs respectively. Nine quintets types (V.1, V.3, V.5, V.7, V.10, V.11, V.13, V.14, and V.16) have 10 MSTs. Two types (V.9 and V.11) have eleven MSTs and other three quintets (V.4, V.15 and V.17) have twelve MSTs.       

The lower value of normalized field $\alpha$ is observed for V.6 and V.8 types with $\alpha=0.5$. That correspond to half the position of the first MSTs of pairs. The largest value of $\alpha$ of the last MSTs is obtained for a group of types (V.2, V.8, V.11, V.12, V.13, V.14, V.15, V.16, V.16, V.17) with $\alpha=25/2$ for individual spin \textit{S} = 5/2. 

\begin{figure}
\centering
\includegraphics[width=8.00 cm]{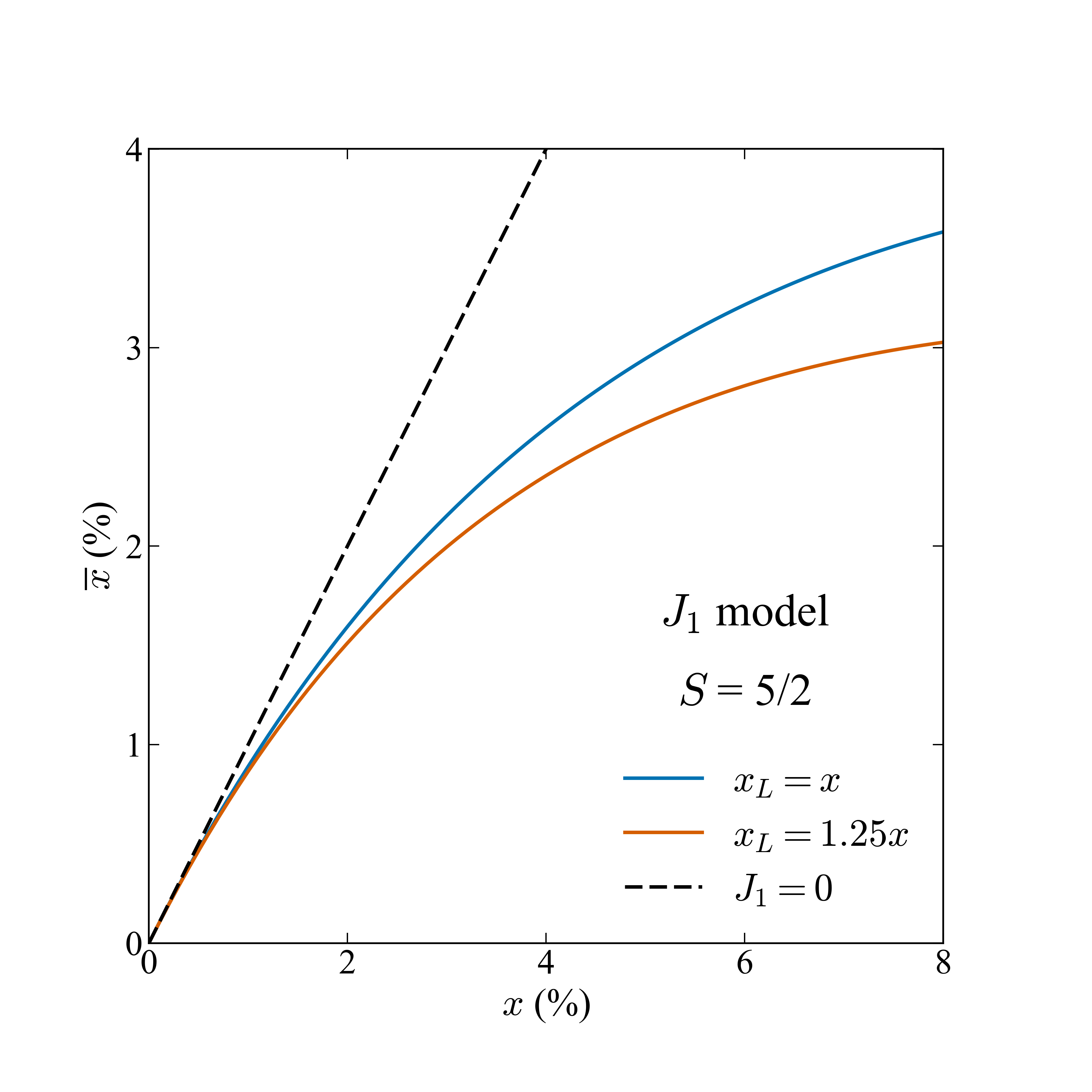}
\caption{Effective concentration $\overline{x}$ as a function of magnetic ions concentration. The solid lines represent the results using Eq.(7) with magnetic ions of spin $S$ = 5/2, for a random distribution ($x_{L} = x$, blue line) and in the case of a clumped distribution with  $x_{L} = 1.25x$ (red line) described in the text. The dashed line represents the result for the paramagnetic case.}
\label{Figure1}
\end{figure}

\begin{table*}[ht]
\centering
\renewcommand{\arraystretch}{2}
\begin{tabular}{|c|c|c|c|}\hline
Type & S & $S_{T}(0) $ & $\alpha = g \mu_{B}H /|2 J/k_{B}|$  \\ \hline
V.1      &\makecell{1/2 \\ 3/2 \\ 5/2} & \makecell{1/2 \\ 3/2 \\ 5/2}  &  \makecell{1.11887, 1.80902 \\ 1.71340, 2.34016, 3.71214, 4.52729, 5.42705 \\ 2.30830, 2.94460, 3.58860, 4.24684, 4.92875, 5.64653, 6.41443, 7.24110, 8.12243, 9.04508}     \\ \hline
V.2      &\makecell{1/2 \\ 3/2 \\ 5/2} & \makecell{3/2 \\ 9/2 \\ 15/2} & \makecell{5/2 \\ 11/2, 13/2, 15/2 \\ 17/2, 19/2, 21/2, 23/2, 25/2 } \\ \hline
V.3      & \makecell{1/2 \\ 3/2 \\ 5/2} & \makecell{1/2 \\ 3/2 \\ 5/2} & \makecell{0.76959, 2.08504 \\ 0.79965, 1.50855, 2.50555,4.08740, 5.19659, 6.25513 \\ 1.08302, 1.60282, 2.31478, 3.22908, 4.35018, 5.99760, 7.19200, 8.29998, 9.37193, 10.42522}    \\ \hline
V.4      & \makecell{1/2 \\ 3/2 \\ 5/2 } &\makecell{1/2 \\ 1/2 \\ 1/2 } & \makecell{0.91496, 2.08504\\ 0.92625, 1.31761, 1.91384, 2.50555, 4.08740, 5.19659, 6.25513 \\ 0.81779, 1.34988, 1.87649, 2.34323, 3.03930, 3.23657, 4.35018, 5.99760, 7.19200, 8.29998, 9.37193, 10.45216}   \\ \hline
V.5      & \makecell{1/2 \\ 3/2 \\ 5/2} & \makecell{1/2 \\ 3/2 \\ 5/2}  & \makecell{0.96665, 2.15139\\ 0.76527, 1.97045, 3.14278, 4.29295, 5.38698, 6.45416 \\ 0.59885, 1.77133, 2.97088, 4.14796, 5.30114, 6.43188, 7.53752, 8.62362, 9.69551, 10.75694}    \\ \hline
V.6      & \makecell{1/2 \\ 3/2 \\ 5/2} &  \makecell{1/2 \\ 5/2 \\ 9/2}  & \makecell{1/2, 5/2\\1/2, 3/2, 11/5, 13/2, 15/2 \\ 1/2, 3/2, 5/2, 17/2, 19/2, 21/2, 23/2}  \\ \hline
V.7      &  \makecell{1/2 \\ 3/2\\ 5/2} &  \makecell{1/2 \\ 3/2\\ 5/2}   &  \makecell{1.21101, 2.24060 \\ 1.80839, 2.70533, 3.67137, 4.67973, 5.70006, 6.72179 \\ 2.47276, 3.31470, 4.21379, 5.15758, 6.13410, 7.13275, 8.14418, 9.16189, 10.18215, 11.20299
}      \\\hline
V.8      & \makecell{1/2 \\ 3/2 \\5/2}& \makecell{1/2 \\ 1/2 \\3/2} & \makecell{1/2, 5/2 \\ 1/2, 3/2, 11/2, 13/2, 15/2 \\ 1/2, 3/2, 5/2, 17/2, 19/2, 21/2, 23/2, 25/2}   \\ \hline
V.9      & \makecell{1/2 \\ 3/2 \\5/2}& \makecell{1/2 \\ 1/2 \\3/2} &\makecell{1.27969, 2.30902\\ 0.60061, 1.76362, 2.80738, 3.84188, 4.87307, 5.90143, 6.92705 \\ 1.13525, 2.24421, 3.29756, 4.33774, 5.37308, 6.40590, 7.43688, 8.46623, 9.49403, 10.52029, 11.54508}    \\ \hline
V.10     & \makecell{1/2 \\ 3/2\\ 5/2}& \makecell{1/2 \\ 3/2\\ 5/2}  & \makecell{1.21101, 2.24060\\ 1.80839, 2.70533, 3.67137, 4.67973, 5.70006, 6.721792 \\ 2.47276, 3.31470, 4.21379, 5.15758, 6.13410, 7.13275, 8.14418, 9.16189, 10.18215, 11.20299
}   \\ \hline
V.11     &  \makecell{1/2 \\ 3/2 \\5/2 }&  \makecell{1/2 \\ 3/2 \\5/2 }  &  \makecell{1, 5/2\\1, 2, 3, 11/2, 13/2, 15/2 \\ 1, 2, 3, 4, 5, 17/2, 19/2, 21/2, 23/2, 25/2}    \\ \hline
V.12     &  \makecell{1/2 \\ 3/2 \\ 5/2}&  \makecell{1/2 \\ 1/2 \\ 3/2} & \makecell{1.207107, 5/2 \\ 0.54370, 1.55342, 2.58678, 3.62132, 11/2, 13/2, 15/2 \\ 
0.93649, 1.92442, 2.94159, 3.97037, 5.00281, 6.03553, 17/2, 19/2, 21/5, 23/5, 25/2} \\  \hline
V.13     & \makecell{1/2 \\ 3/2 \\ 5/2}& \makecell{1/2 \\ 3/2 \\ 5/2} &  \makecell{1, 5/2 \\1, 2, 3, 11/2, 13/5, 15/2\\ 1, 2, 3, 4, 5, 17/2, 19/2, 21/2, 23/2, 25/2}   \\ \hline
V.14     &\makecell{1/2 \\ 3/2 \\ 5/2}&\makecell{1/2 \\ 3/2 \\ 5/2}  & \makecell{3/2, 5/2 \\ 5/2, 7/2, 9/2, 11/2, 13/2, 15/2 \\ 7/2, 9/2, 11/2, 13/2, 15/2, 17/2, 19/2, 21/2, 23/2, 25/2}  \\ \hline
V.15     & \makecell{1/2 \\ 3/2 \\ 5/2}& \makecell{1/2 \\ 1/2 \\ 1/2} & \makecell{3/2, 5/2 \\ 3/2, 5/2, 7/2, 9/2, 11/2, 13/2, 15/2 \\ 3/2, 5/2, 7/2, 9/2, 11/2, 13/2, 15/2, 17/2, 19/2, 21/2, 23/2, 25/2} \\ \hline
V.16     &\makecell{1/2 \\ 3/2 \\ 5/2} &\makecell{1/2 \\ 3/2 \\ 5/2} &  \makecell{3/2, 5/2\\ 5/2, 7/2, 9/2, 11/2, 13/2, 15/2 \\ 7/2, 9/2,    11/2, 13/2, 15/2, 17/2, 19/2, 21/2, 23/2, 25/2}   \\ \hline
V.17     & \makecell{1/2 \\ 3/2 \\ 5/2 }&\makecell{1/2 \\ 1/2 \\ 1/2 } &  \makecell{3/2, 5/2 \\3/2, 5/2, 7/2, 9/2, 11/2, 13/2, 15/2 \\ 3/2, 5/2, 7/2, 9/2, 11/2, 13/2, 15/2, 17/2, 19/2, 21/2, 23/2, 25/2}  \\ 
 \hline
\end{tabular}
\caption{Normalized fields at the MSTs for the seventeen quintet types in a model with one exchange constant between the twelve nearest neighbors of the hcp lattice. The fields at the steps are given by $H = 2\alpha |J/k_{\mathrm{B}}|/g \mu_{B}$. The results are for clusters with identical magnetic ions with spin \textit{S} = 1/2, 3/2, and 5/2. $S_{T}(0)$ is ground state total spin at zero magnetic field.}
\label{Table1}
\end{table*}

\section{IV. Technical saturation}

Technical saturation corresponds to the saturation of the magnetization of clusters with ground state total spin at zero field, $S_{T}(0)$, different from zero. The $J_{1}$ model includes the contributions of singles, pairs, two types of triplets (open and closed triplets), six types of quartets, and seventeen types of quintets. The singles, open triplets with $S_{T}(0) = S$, closed triplets with $S_{T}(0)$ = 1/2, propeller quartets (PQ) with $S_{T}(0) = 2S$ and funnel quartets (FQ) with $S_{T}(0)$ = (2\textit{S}-1)/2 contribute to the technical magnetization. As we can see in Table I, all the quintet types have non zero $S_{T}(0)$ and thus contribute to the technical saturation. The ratio of the technical saturation value over the total saturation $M_{S}/M_{0}$ is the sum of the contributions from clusters with $S_{T}(0) \neq 0$. Each contribution depends on the probability of the cluster type and the ratio between $S_{T}(0)$ and the total spin $S_{T}$ = $S$ $\times$\textit{i} where \textit{i} is the number of atoms in the cluster. The probability that a spin belongs to singles, open triplets, closed triplets is $P_{S} = (1-x)^{12}$, $P_{OT} = 18x^{2}(1-x)^{23}(2+5(1-x)) $, and $P_{CT} = 3x^{2}(1-x)^{21}(1+6(1-x)+(1-x)^{2})$ respectively. For the funnel quartets, we have $P_{FQ} =24x^{3}(1-x)^{27}(8+11(1-x)+(1-x)^{2})$ and for the propeller quartets $P_{PQ} = 16x^{3}(1-x)^{28}(3+6(1-x)+2(1-x)^{2})$. The probability that a spin is within a quintet type is given in Table II. The expressions have been obtained using the cluster tables \cite{ClusterTables} based  on a random distribution of the magnetic ions over the cation sites of the lattice.

The ratio $M_{S}/M_{0}$ (or the ratio $\overline{x}/x$ of the effective concentration to the total concentration) is given by: 

\begin{equation} \label{eq1}
\begin{split}
M_{S}/M_{0}  = P_{S} + P_{OT}/3 + \beta P_{CT} + P_{PQ}/2 + \gamma P_{FQ}\\
+\sum  \limits_{i=1}^{17} F_{V.i}P_{V.i} + P_{>5}/7 
\end{split}
\end{equation}
where the factors $\beta$ = 1/3, 1/9, and 1/15 and $\gamma$ = 0, 1/6, 1/5 for quartets of individual spin \textit{S} = 1/2, 3/2 and 5/2, respectively. $P_{V.i}$ is the concentration of each quintet type V.i (Table II). The factors $F_{V.i}$ are given by $S_{T}(0)$/(5\textit{S}) using the values of $S_{T}(0)$ and \textit{S} of Table I. The last term in Eq.(7) represents the contribution of the clusters larger than quintets. $P_{>5}$ is the probability of the cluster larger than $n$ = 5. All the clusters with $n$ $>$ 5 are assigned to septet (\textit{n} = 7) \textit{strings} with $S_{T}(0)$/(7\textit{S})= 1/7. Figure 4 shows the effective concentration $\overline{x}$ as function of the total magnetic ion concentration obtained using Eq.(7). The dashed line represents the paramagnetic case. As an example of the application of Eq.(7),  consider a spherical particle of DMS with a core free of magnetic ions and a shell  enriched with magnetic ions. In the shell the distribution of the magnetic ions is assumed to be random. The ratio of the volume of the shell ($V_{Shell}$) to the volume of the core ($V_{Core}$) is given by $V_{Shell}$/$V_{Core}$ = $\delta$.  The concentration of the magnetic ions is calculated using the total volume of the particle ($V_{Shell}$+$V_{Core}$), however, the magnetic properties and thus the effective concentration are determined by the volume of the shell. The effective concentration is then given by Eq.(7) using the local concentration $x_{L} = (1+\delta)x/\delta$ instead of $x$. Figure 4 shows the effective concentration as function of the concentration of the magnetic ions calculated by Eq.(7), in the case of a random distribution ($x = x_{L}$) and in the case of the spherical particle described above with $V_{Shell}$/$V_{Core}$ = 4.

\section{V. Conclusion}

We have calculated the magnetization of quintets in the model with AF exchange interaction between the 12 nearest neighbors in the wurtzite structure. The choice $J_{1}^{in} = J_{1}^{out}$ was made to reduce the number of quintet types and simplify the calculation. The fields at the MSTs have been calculated by diagonalizing the spin Hamiltonian matrix in the basis of states which allow the incorporation of single ion anisotropy. The updated expression for the ratio $M_{S}/M_{0}$ can be used in the study of the distribution of the magnetic ions in DMSs nanoparticles.   

\begin{center}
%$\mathbf(Acknowledgments)$
$Acknowledgments$
\end{center}
This work was supported by FAPESP (Fundação de Amparo à Pesquisa do Estado de São Paulo, Brazil) under Contract No. 2015/16191-5 and by CNPq (Conselho Nacional
de Desenvolvimento Científico e Tecnológico, Brazil) under Contract 306715/2018-0.
\vspace{-0.4cm}

\begin{table*}[ht]
\centering
\renewcommand{\arraystretch}{1.5}
\begin{tabular}{|c|c|}\hline
Quintet type& Probability  \\ \hline
V.1      & $10x^{4}(1-x)^{33}(24+153(1-x)+414(1-x)^{2}+360(1-x)^{3})$       \\ 
 \hline
V.2      &$15x^{4}(1-x)^{32}(1+2(1-x)^2)$       \\ \hline
V.3      &$60x^{4}(1-x)^{33}(2+23(1-x)+34(1-x)^{2}+10(1-x)^{3})$      \\ \hline
V.4      &$10x^{4}(1-x)^{32}(30+180(1-x)+171(1-x)^{2}+15(1-x)^{3})$    \\ \hline
V.5      &$30x^{4}(1-x)^{32}(9+51(1-x)+35(1-x)^{2}+2(1-x)^{3})$       \\ \hline
V.6      &$45x^{4}(1-x)^{32}(4+7(1-x))$       \\ \hline
V.7      &$120x^{4}(1-x)^{31}(1+(1-x))$       \\ \hline
V.8      &$30x^{4}(1-x)^{31}(7+5(1-x))$       \\ \hline
V.9      &$30x^{4}(1-x)^{30}(3+(1-x))$       \\ \hline
V.10     &$30x^{4}(1-x)^{30}(2+19(1-x)+18(1-x)^{2})$       \\ \hline
V.11     &$30x^{4}(1-x)^{30}(3+16(1-x)+7(1-x)^{2})$       \\ \hline
V.12     &$15x^{4}(1-x)^{29}(7+18(1-x)+(1-x)^{2})$       \\ \hline
V.13     &$30x^{4}(1-x)^{30}(2+3(1-x))$       \\ \hline
V.14     &$15x^{4}(1-x)^{29}$       \\ \hline
V.15     &$30x^{4}(1-x)^{28}$       \\ \hline
V.16     &$30x^{4}(1-x)^{28}(1+2(1-x))$       \\ \hline
V.17     &$5x^{4}(1-x)^{27}$       \\
 \hline
\end{tabular}
\caption{Probability that a spin belongs to one of the seventeen quintet types in the isotropic $J_{1}$ model for the hcp lattice. The expression have been synthesized using the cluster tables \cite{ClusterTables}.}
\label{Table1}
\end{table*}

\bibliographystyle{apsrev4-2}
\bibliography{Report}

\begin{thebibliography}{11}
\bibitem [{Clu()}]{ClusterTables}%
  \BibitemOpen
  \href@noop {} {\bibinfo {title} {The cluster tables can be found in the
  suplementary files of {Ref}. \cite{10.1063/1.1507808}}}\BibitemShut {NoStop}%
\end{thebibliography}

\end{document}